\documentclass[manuscript,screen]{acmart}

\AtBeginDocument{%
  }

\setcopyright{acmlicensed}
\copyrightyear{2018}
\acmYear{2018}
\acmDOI{XXXXXXX.XXXXXXX}

\acmConference[Conference acronym 'XX]{Make sure to enter the correct
  conference title from your rights confirmation email}{June 03--05,
  2018}{Woodstock, NY}

\setcitestyle{authoryear}

\usepackage{subcaption}
\usepackage[dvipsnames]{xcolor}
\usepackage{hyperref}
\usepackage{flafter}
\usepackage{graphicx}
\setkeys{Gin}{draft=false}

\begin{document}

\title{Hyper--learning and Unlearning: A Narrative Speculation on Urbanism in Media Ecologies}

\author{Anqi Wang}
\email{awangan@connect.ust.hk}
\affiliation{%
  \institution{HKUST}
  \country{Hong Kong SAR}}
\affiliation{%
  \institution{University College London}
  \city{London}
  \country{UK}}

\author{Yue Hua}
\affiliation{%
\institution{Shanghai Jiaotong University}
  \city{Shanghai}
  \country{China}}
\affiliation{%
  \institution{University College London}
  \city{London}
  \country{UK}}
\email{yue.hua.20@alumni.ucl.ac.uk}

\author{Xinyue Zhang}
\affiliation{%
  \institution{University College London}
  \city{London}
  \country{UK}
}
  \email{xinyue_zhang.20@alumni.ucl.ac.uk}

\author{Jindi Jia}
\affiliation{%
  \institution{University College London}
  \city{London}
  \country{UK}
}
\email{jindijia97@gmail.com}

\author{Corneel Cannaerts}
\affiliation{%
  \institution{KU Leuven}
  \city{Brussels}
  \country{Belgium}
}
\affiliation{%
  \institution{University College London}
  \city{London}
  \country{UK}}
\email{corneel.cannaerts@kuleuven.be}

\author{Michiel Helbig}
\affiliation{%
 \institution{KU Leuven}
 \city{Brussels}
 \country{Belgium}
 }
 \affiliation{%
  \institution{University College London}
  \city{London}
  \country{UK}}
\email{michiel.helbig@kuleuven.be}

\author{Pan Hui}
\affiliation{%
  \institution{HKUST (Guangzhou)}
  \city{Guangzhou}
  \country{China}}
\affiliation{%
  \institution{HKUST}
  \country{Hong Kong SAR}}
\email{panhui@ust.hk}

\renewcommand{\shortauthors}{Anqi Wang et al.}



\received{20 February 2007}
\received[revised]{12 March 2009}
\received[accepted]{5 June 2009}

    \begin{teaserfigure}
        \centering 
        \includegraphics[width=\textwidth]{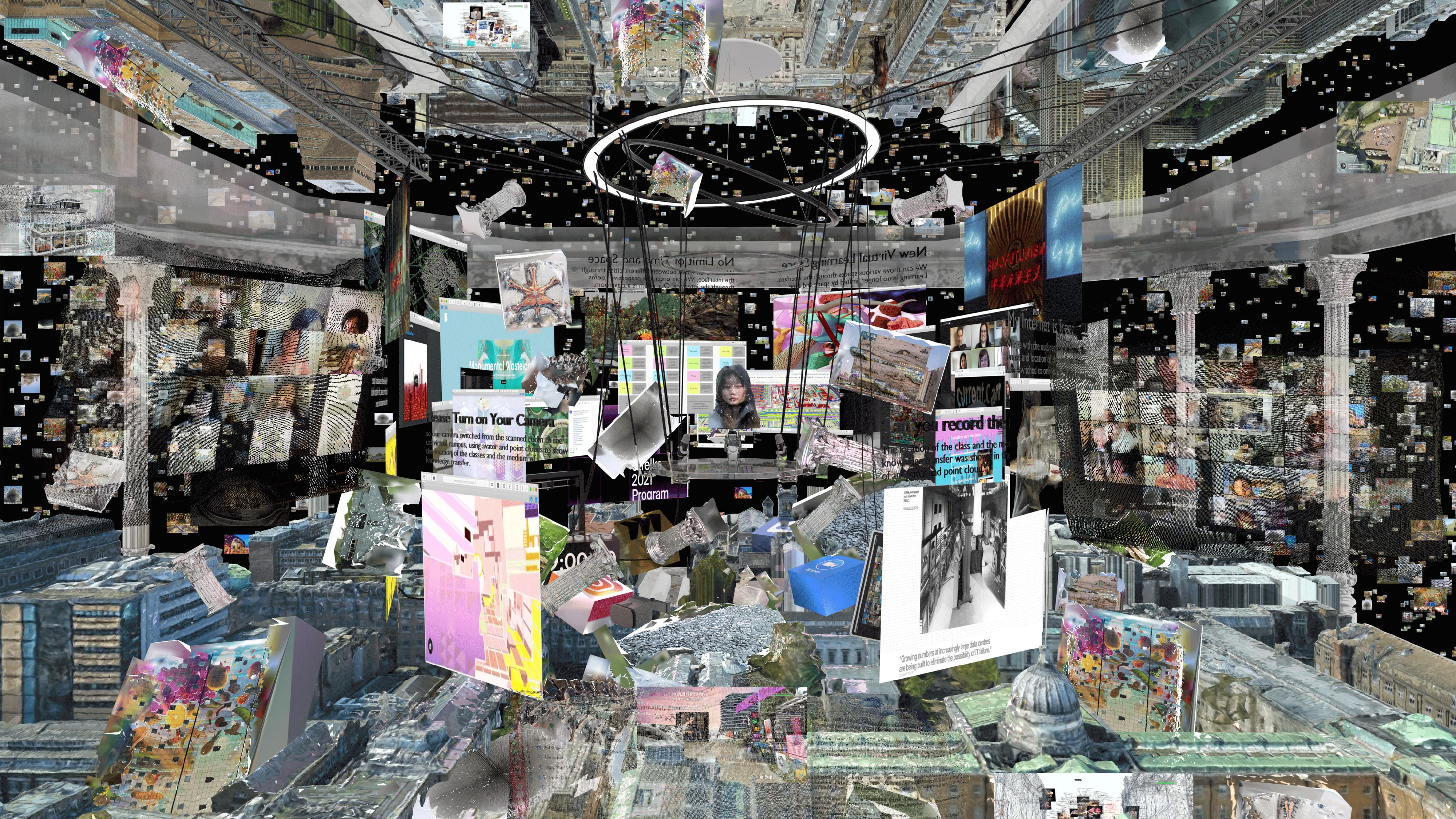} 
        \caption{A speculative XR learning environment in which urban ruins, photogrammetric city fragments, and platform interfaces collapse into a single compressed spatial field. Floating screens, point clouds, and archival media assemble into an unstable ``urban learning space,'' visualizing hyper-learning and unlearning as distributed, algorithmically mediated processes embedded within the city's media ecology.}
        \label{fig:teaser} 
    \end{teaserfigure}

\begin{abstract}
    \emph{Hyper--Learning and Unlearning} is a speculative animation that reflect how learning is reconfigured within digital media ecologies. Using architectural education as a microcosm, the work reframes the city as a hyper-learning apparatus where urban space, algorithmic systems, and platform infrastructures condition cognition and agency. By staging both hyper-learning and the unlearning induced by machine-supported cognition, the work critiques institutional gatekeeping while revealing how platforms reshape expertise, memory, and spatial experience. This project invites viewers to reconsider how urban space becomes pedagogical infrastructure in a posthumanism era. 
\end{abstract}

\maketitle

\section{Introduction}
    \emph{Hyper-Learning \& Unlearning} (Figure~\ref{fig:teaser}) is a speculative extended-reality (XR) art project that interrogates how digital media ecologies~\cite{fuller2005mediaecologies,cinque2024medialogy,taffel2019digital} are reshaping the conditions under which knowledge is produced, transmitted, and embodied~\cite{posthuman2013braidotti,mcluhan1964understanding,Braidotti2016}. As digital platforms increasingly permeate urban and educational life, the project interrogates how learning is transformed when information becomes ubiquitous, attention is fragmented across layered infrastructures, and the city itself operates as a condensed interface of physical, virtual, and algorithmic forces~\cite{bratton2015,kitchin2014real}. 
~
Positioning architecture as a posthuman learning infrastructure emerging after the \emph{Digital Second Turn}~\cite{carpo2017second}, \emph{Hyper-Learning \& Unlearning} reconceptualizes space not as a passive container of education but as an active system that algorithmically produces subjectivity through media saturation, platform logics, and infrastructural compression~\cite{foucault1984space}. The city is thus treated as a distributed interface—one that continuously learns, forgets, and recomposes subjects through spatialized computation~\cite{latour2005reassembling,suchman2007humanmachine}.

Building on Marshall McLuhan’s proposition that media extend and reconfigure human capacities~\cite{mcluhan1964understanding}, digital media ecologies have evolved into dense assemblages of platforms, sensors, databases, and machine-learning systems that increasingly learn about subjects as much as subjects learn through them. As media ecology undergoes a posthumanist turn, it not only reveals how technological environments shape perception but also raises a critical question: whether the notion of the \textit{``subject''} itself is an effect of these systems~\cite{hayles1999,bratton2015}. Drawing on posthumanist frameworks~\cite{forlano2017posthuman,posthuman2013braidotti}, this project reconceptualizes subjectivity not as a pre-given condition but as one \textit{constituted} through educational and technological infrastructures.

By embedding learning within a network of hybrid spaces—physical streets, virtual classrooms, and machine-curated archives—the project renders visible the compression of educational environments into algorithmically modulated flows of images, tasks, and behaviors. It foregrounds a future in which pedagogical authority is redistributed among humans, machines, and urban infrastructures, raising critical questions about which forms of knowledge must be learned, unlearned, or delegated~\cite{latour2005reassembling,forlano2017posthuman}. Through this immersive and speculative reconfiguration of architectural education, the project invites viewers to reconsider the role of media technologies in shaping knowledge production, agency, and participation within increasingly compressed urban and cognitive landscapes.
~
This study translates theoretical debates spanning media ecology and posthumanism into artistic practice through material and computational engagement. Contributing to computational art and speculative design, it demonstrates how point cloud rendering~\cite{ivsic2022artpoint}, photogrammetric scraping~\cite{sholarin2015photogrammetry}, and XR game environments operate as forms of media-archaeological methodology~\cite{huhtamo2011,parikka2012}. By framing learning systems as spatially distributed and algorithmically generative fields of posthuman subjectivity, the project advances an aesthetic and critical lens for understanding the compression dynamics of networked society. 

\section{Related Work}
    \subsection{Transformative City in Media Ecology}

Media ecologies force to reorganize perception, cognition, and social relations of city~\cite{mcluhan1964understanding,postman1970,fuller2005mediaecologies,cinque2024medialogy,taffel2019digital}. Urban space thus features a mediated milieu in which learning, authority, and knowledge circulation are continuously restructured~\cite{meyrowitz1985}. With the rise of digital networks, scholars argue that cities undergo a simultaneous collapse of distance and intensification of place, producing hybrid spatialities governed by informational flows rather than fixed geographies~\cite{castells1996network,graham1998end}.  

Media archaeology deepens this account by tracing how historical urban and educational media persist as recompressed residues within contemporary technical systems~\cite{huhtamo2011,parikka2012}. Rather than disappearing, architectural pedagogies, diagrams, and institutional spaces are decomposed and reassembled as databases, interfaces, and computational images~\cite{manovich2002,manovich2013}. This recompression aligns with broader theories of time--space condensation, where acceleration and abstraction destabilize embodied learning and spatial memory~\cite{harvey1989}.  




\subsection{Speculation as a Narrative Urbanism}
Speculative imagination frames future-making as a narrative and critical practice rather than a predictive exercise~\cite{slaughter1998dystopia}. Within design research, speculative design has developed as a means of interrogating dominant technological agendas by foregrounding what remains absent, marginalized, or unarticulated~\cite{lindley2015back,Bratteteig2012PD}. Such futuring practices operate as critical interventions that can reconfigure existing social orders through narrative displacement and imaginative reconstruction~\cite{Sönmez2023}. 

In urban discourse, speculation functions as narrative urbanism, where storytelling and computational mediation become spatial practices~\cite{corneel2025entangled}. For example, narrative that is fragmented and staged to surface latent political and infrastructural assumptions~\cite{corneel2025entangled,banham1960megstructure}. Liam Young employ cinematic storytelling and fictional urban worlds as critical instruments for interrogating planetary-scale urbanization~\cite{young2021planetcity}. 
Recent Siggraph art practices extend this lineage by operationalizing speculative narratives through computational media, immersive environments, and performative systems. Works such as \textit{Drift of the Uncharted}~\cite{ary2025drift}, \textit{Water City}~\cite{yoshida2024speculative}, \textit{Abutting Tokyo}~\cite{zhong2024abutting}, and \textit{Algorithmic Miner}~\cite{sun2025algorithmic} employ robotics, generative imagery, and virtual reality to render climate change, data bias, and hidden labor as experiential urban narratives.

\section{Conceptual Framework: A Speculative Narrative}
Building on these reflection on theories and social phenomenons, we explore the concept framework of \emph{Hyper-learning and Unlearning} in positioning (Section~\ref{sec:position}) and narrative (Section~\ref{sec:narrative}) aspects. 

\begin{figure*}[h]
    \centering
    \includegraphics[width=\linewidth]{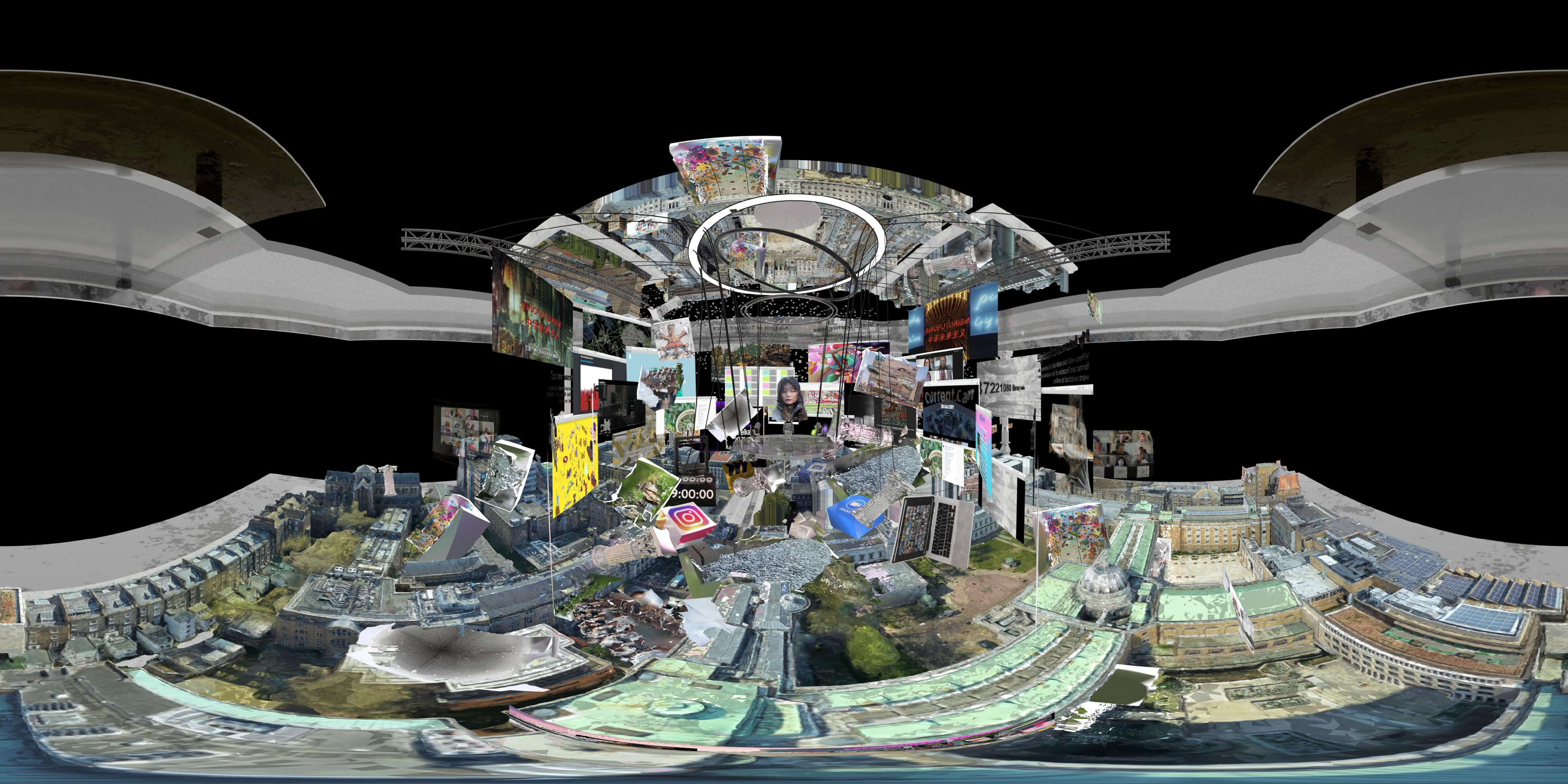}
    \caption{The central avatar operates as a distributed interface, mediating and connecting multiple urban environments.}
    \label{fig:teaser2-avatar}
\end{figure*}


\subsection{Positioning}~\label{sec:position}
This approach builds on recent intersections of digital media and design research:
\textbf{STS and Platform Studies}: Bogost and Montfort's ``platform studies''~\cite{MontfortBogost2009PlatformStudies} demonstrate how technical constraints shape expressive possibilities; we extend this to pedagogical platforms, treating gamification and XR as historically contingent rather than inevitable infrastructures.
\textbf{Computational Aesthetics}: Manovich's analysis of ``software takes command'' \cite{manovich2013} provides a framework for understanding how algorithmic operations become aesthetic conventions; our point-cloud rendering makes these operations visible as media-archaeological evidence.
\textbf{Critical Making}: Ratto's ``critical making''~\cite{Ratto2011CriticalMaking} and Sayers' work on ``prototyping the past''~\cite{Sayers2015Prototyping} inform our use of 3D scanning and photogrammetry~\cite{sholarin2015photogrammetry} not as neutral documentation tools but as archaeologically inflected making practices that generate critical insight through technical manipulation.

\subsection{Speculative Narrative Apparatus}~\label{sec:narrative} 
Grounding on the positioning, we crafted the core experimental narrative of this project (Figure~\ref{fig:teaser2-avatar}). The narrative structure comprise of two parts, as shown in Figure~\ref{fig:3.2.1} and Figure~\ref{fig:narrative}. 


\subsubsection{Metaphor: Pedagogical Ruins--Media Archaeology in Point Clouds}~\label{sec:3.2.1}
\begin{figure}[h]
    \centering
    \includegraphics[width=0.78\linewidth]{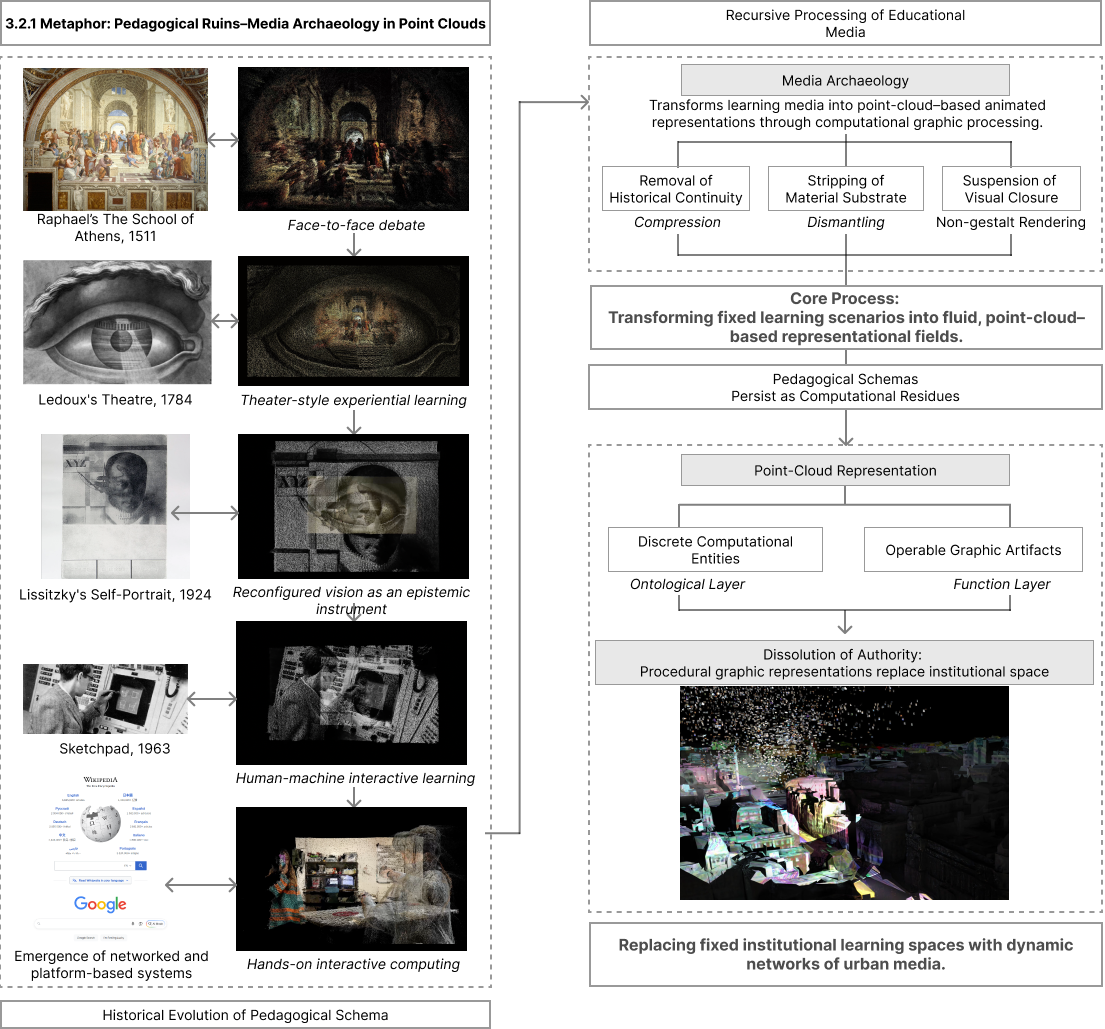}
    \caption{
    Conceptual framework illustrating the \emph{Metaphor} part of the animation structure, in which learning milestones are translated into point-cloud representations through a media-archaeological lens.}
    \label{fig:3.2.1}
\end{figure}
Canonical pedagogical schema has not vanished, but persists in the form of ruins—constantly dismantled, compressed, and reassembled by technology. As Figure~\ref{fig:3.2.1}, the animation's successive dissolving layers of 3D point cloud suggest this transformation: the \textit{School of Athens}~\cite{Raphael_SchoolOfAthens_1511} staged knowledge as embodied dialogue within a shared architectural frame. Ledoux’s theater~\cite{Ledoux_TheatreBesancon_1784} formalized pedagogy as collective attention disciplined by spatial design. Lissitzky’s self-portrait~\cite{Lissitzky_Constructor_1924} reconfigured vision itself as an epistemic instrument, anticipating modernist beliefs in human–machine alignment. Sketchpad
    ~\footnote{It pioneered human–computer interaction (HCI), and is considered the ancestor of modern computer-aided design (CAD) programs and as a major breakthrough in the development of computer graphics in general. Source:~\url{https://en.wikipedia.org/wiki/Sketchpad}}
~\cite{BiMplus_Sketchpad_Article,sutherland1963_sketchpad, sketchpad1998ivane} marked a decisive shift toward interactive computation, where drawing and reasoning converged through direct manipulation \cite{Myers_HCI_1998}. Amidst these overlapping screenage, these forms are rendered unstable, their didactic logics reduced to data points that no longer sustain institutional authority.

The subsequent emergence of networked and platform-based systems—the World Wide Web as hypertextual infrastructure \cite{BernersLee_WWW_1989}, algorithmic search as epistemic interface \cite{Brin_PageRank_1998}, multi-touch mobile devices \cite{Han_MultiTouch_2005}, and online learning environments that exceed the classroom \cite{Moore_DistanceEducation_2011,Kizilcec_MOOCs_2017}--is not framed as technological advancement, but as a process of pedagogical compression. These systems accelerate access to knowledge while simultaneously dissolving the spatial, temporal, and institutional frames that once stabilized learning practices.

~~~~
The point-cloud aesthetic operates as a form of media-ecological archaeology \cite{parikka2012}. By leveraging the discreteness and transience of point-based representation, the work extracts pedagogical media from their historically continuous and stable material substrates, reconstructing a cognitive trajectory that extends from Renaissance idealized spaces to contemporary XR interfaces~\cite{manovich2002}. What is rendered is not a seamless visual object, but a field of computational fragments. This mode of representation aligns with what Grosoli terms \textit{``non-gestalt rendering,''}~\cite{} in which images refuse immediate perceptual closure and instead remain partially unresolved, contingent, and computationally legible. Such refusal interrupts the image’s function as a terminal site of meaning and repositions it as an operable artifact \cite{manovich2013}—one that can be decomposed, recomputed, and reorganized within technical systems.

This disassembly of pedagogical form resonates with Vilém Flusser’s account of the recursive nature of technical images, whereby new media emerge through the ingestion and reprogramming of prior representational regimes \cite{flusser2011}. When the architectural arcades of \textit{The School of Athens} collapse into algorithmically generated point clouds, the work exposes a critical metaphorical shift: educational ideals grounded in embodied co-presence and spatial continuity are reconfigured as distributed, transmissible data flows within networked interfaces. This transformation signals a broader media-historical migration of cultural memory—from continuous narratives to discrete, modular structures \cite{flusser2011,hayles2005,manovich1999}. 
In this sense, the point cloud functions simultaneously as a technical apparatus and a critical metaphor: it endows classical pedagogical images with a digital body while revealing how any pedagogical framework, under conditions of technological iteration, persists only as a provisional and contingent media construct~\cite{kittler1999,stiegler1998}.

\subsubsection{Prologue: Entering an Gamified XR World}~\label{sec:3.2.2}
\begin{figure*}
    \centering
    \includegraphics[width=\linewidth]{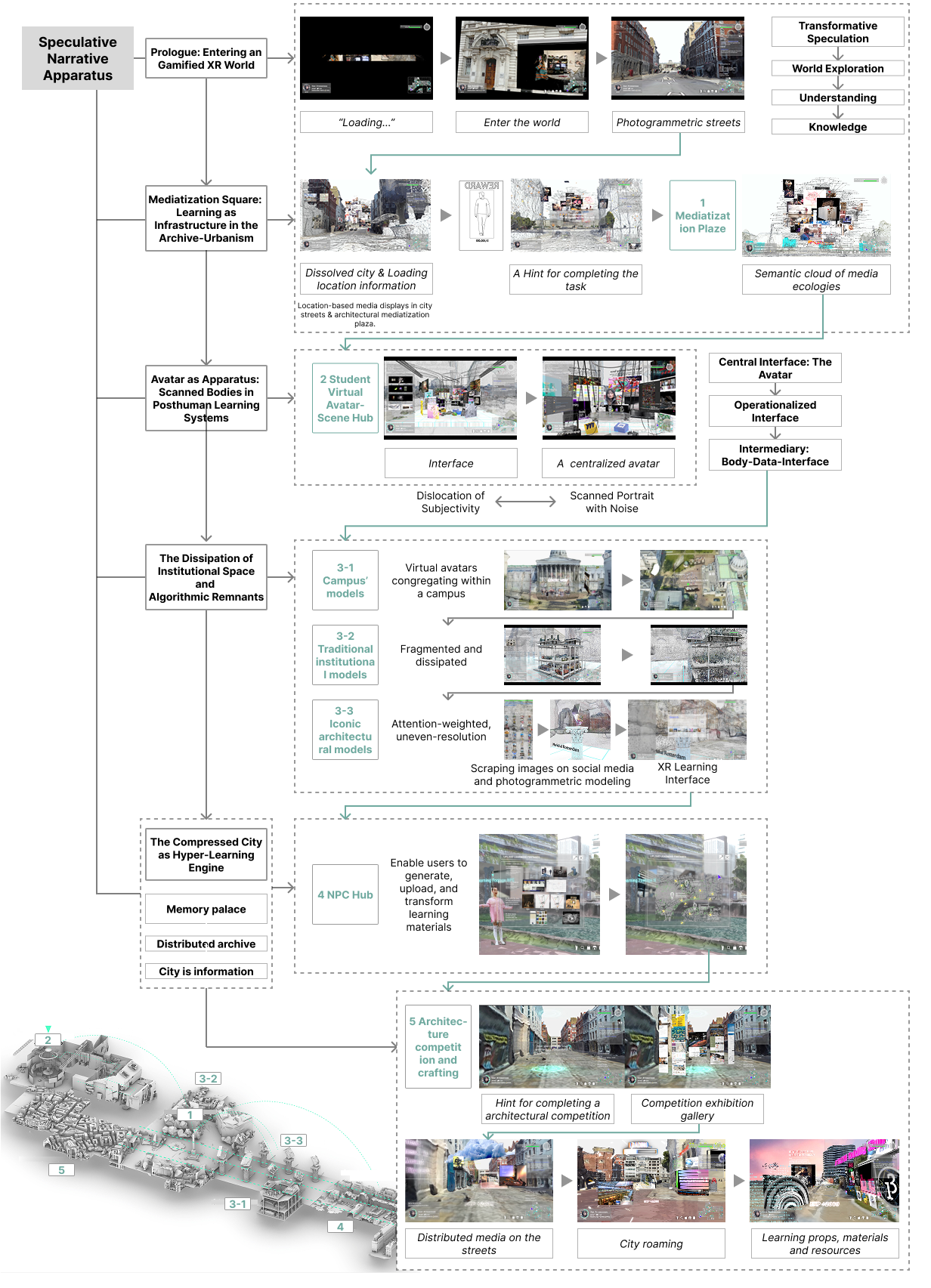}
    \caption{Narrative structure of the speculative XR world experience, outlining the progression of speculative scenarios and conceptual lenses.}
    \label{fig:narrative}
\end{figure*}
The fragility revealed through the visual motif of ``pedagogical ruins'' (Section~\ref{sec:3.2.1}) functions as a ritualized threshold into a gamified mixed-reality city. A gradually unfolding XR system interface signals the reconfiguration of cognitive frameworks within a blurred physical--virtual continuum (Figure~\ref{fig:narrative}). Within this emergent gamified environment, a player avatar named \emph{Arrowanneee} appears, establishing the experiential lens through which the city is navigated. The ensuing urban traversal unfolds from this perspective, positioning gameplay as both a perceptual interface and an epistemic mediator.
This animated sequence signals the entry into a world in which educational forms and technological interfaces can be systematically traced, deconstructed, and reconfigured. Here, the ``ruins'' of historical pedagogy are reframed as productive resources for future creation. 

In this context, the project advances a foundational ontological claim: every pedagogical landscape is constituted by specific technological preconditions~\cite{kittler1999}. These preconditions inevitably surface through media-archaeological processes, exposing the historical and material substrates that structure contemporary knowledge systems~\cite{huhtamo2011,parikka2012,benjamin1999,chun2011}.
This revelation further engages with ontological inquiry by addressing how the XR game world is constructed and governed by its underlying rules, thereby resonating with established perspectives in game ontology. It establishes the philosophical and narrative foundation for subsequent gameplay experiences, in which players actively assemble new learning scenarios from fragments of historical media.


\subsubsection{Mediatization Square:~Learning as Infrastructure in the Archive--Urbanism}
\begin{figure*}[h]
    \centering
    \includegraphics[width=\linewidth]{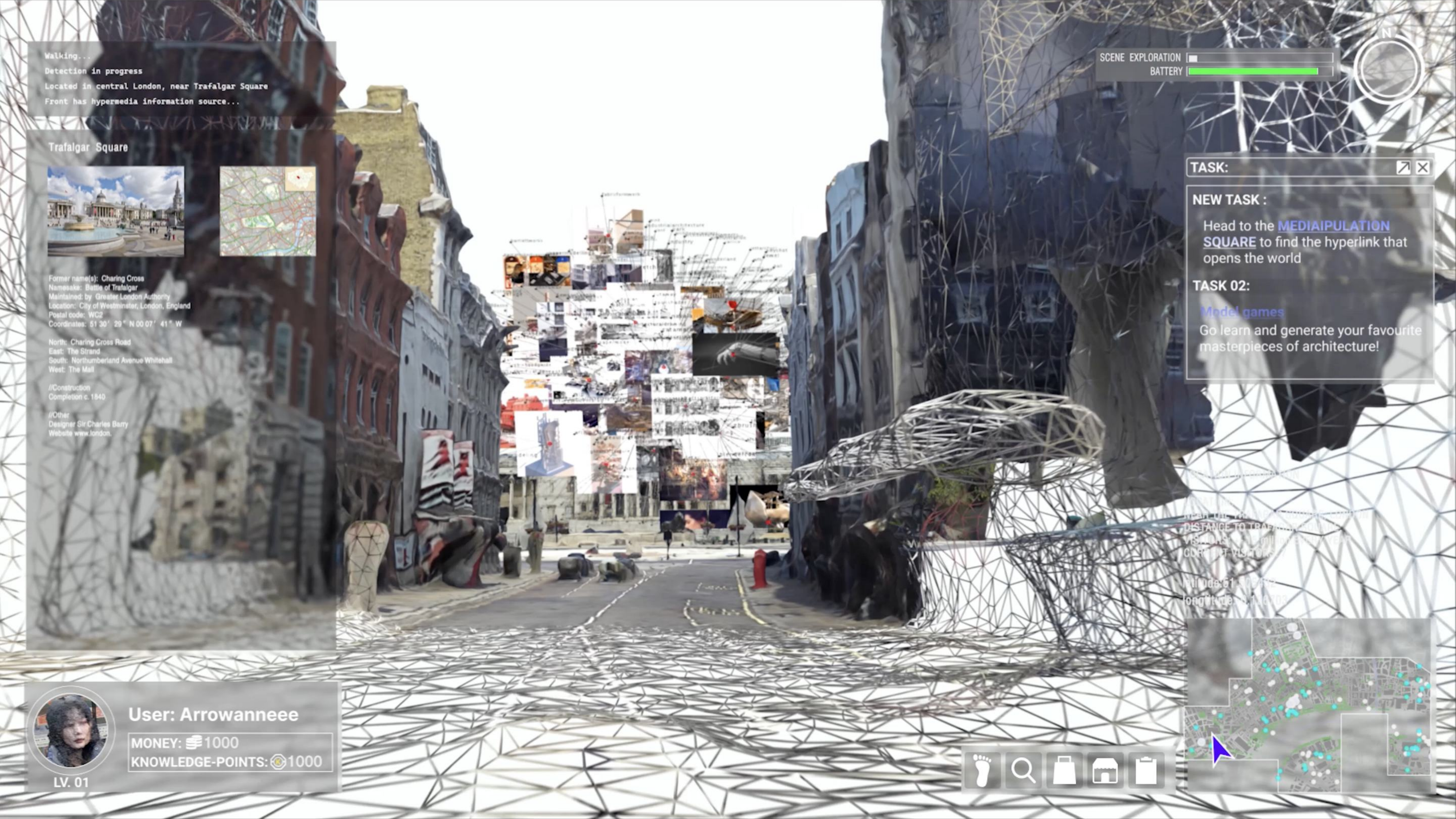}
    \caption{Dissolved city streets rendered as fragmented and reconfigured urban infrastructures.}
    \label{fig:streetpolygan}
\end{figure*}
The player transition from photogrammetrically rendered street scenes into a city that progressively dissolves into grids and point clouds (Figure~\ref{fig:narrative}). As architectural facades are algorithmically stripped away, urban space is no longer perceived as a stable environment but exposed as \textit{``computable residue''}—a substrate shaped by data operations rather than human-scale materiality (Figure~\ref{fig:streetpolygan}).
Simultaneously, heads-up display (HUD) elements—task bars, energy meters, and mini-maps—overlay the city, recoding it as an operable interface. Within this mediated regime, learning is reconfigured as a sequence of calculable actions: \emph{exploration} is reduced to task completion, \emph{understanding} is displaced by mechanisms of collection and unlocking, and \emph{knowledge} is quantified as points, scores, and rewards. Rather than residing within the learner, educational processes are externalized and governed through interface logic and system feedback. 

As the avatar roams, a media archival cluster situated at Trafalgar Square--\emph{``Mediatization Square''} came into view (Figure~\ref{fig:narrative}1). Composed of image assemblages, radiating keywords, and density-based relational links, the cluster forms a \textit{``relational knowledge topography''} that renders distributed, networked media traces spatially legible. Although the underlying data originates from decentralized and heterogeneous sources, its concentration within a historically centralized urban site foregrounds the infrastructural mechanisms through which media ecologies momentarily re-center dispersed information in order to be perceived, navigated, and acted upon.

In this sense, \emph{Mediatization Square} does not function as a locus of authority, but as an apparatus of appearance that exposes how learning, under posthuman conditions, is increasingly produced by archival systems, algorithmic associations, and interface-driven mediation rather than by individual cognition alone~\cite{kittler1999,parikka2012,chun2011}.



\subsubsection{Avatar as Apparatus: Scanned Bodies in Posthuman Learning Systems}
\begin{figure*}[h]
    \centering
    \begin{subfigure}[t]{\textwidth}
        \centering
        \includegraphics[height=8cm]{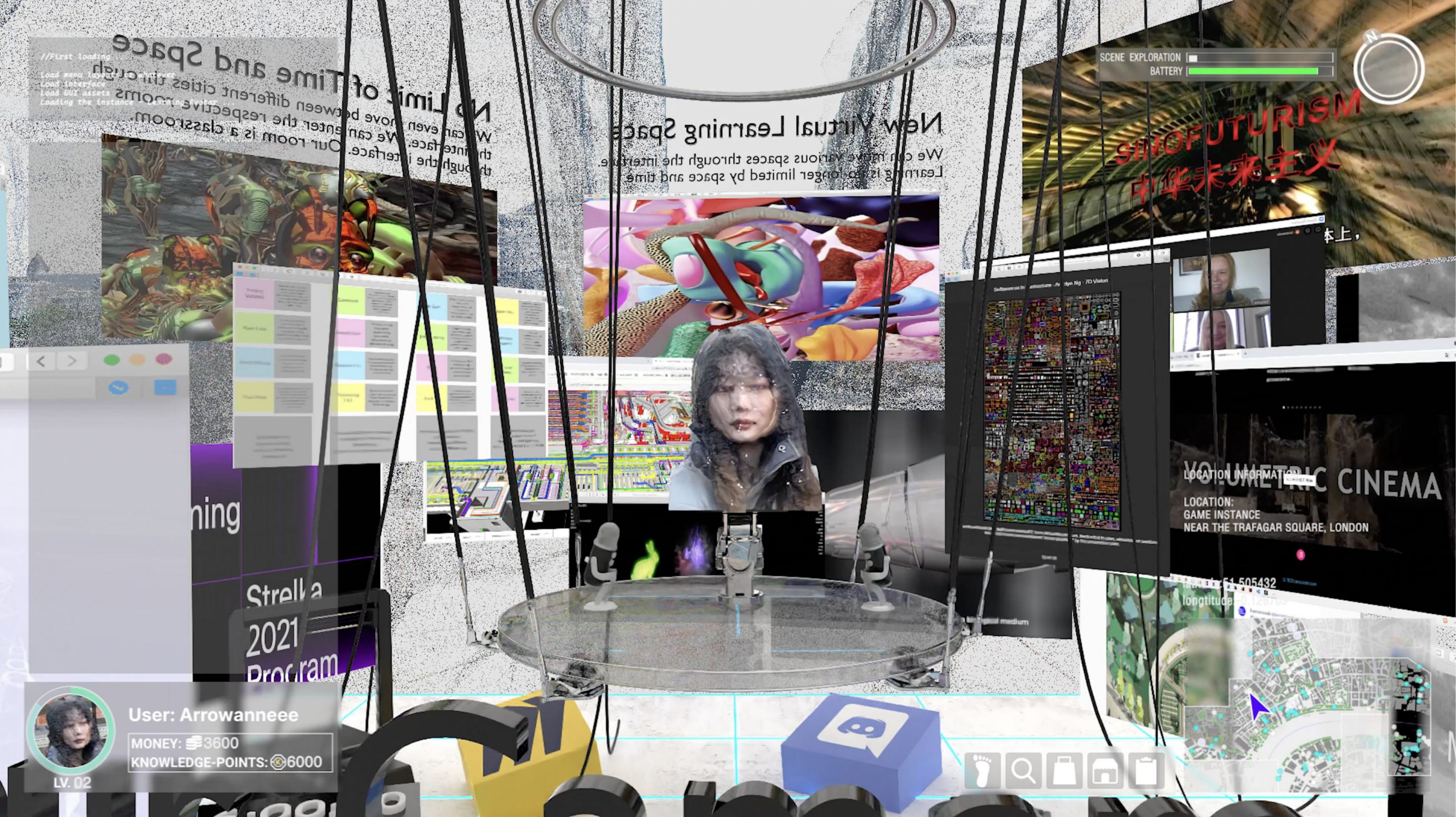}
        \label{fig:avatar1}
    \end{subfigure}
    \vspace{2mm}
    \begin{subfigure}[t]{\textwidth}
        \centering
        \includegraphics[height=8cm]{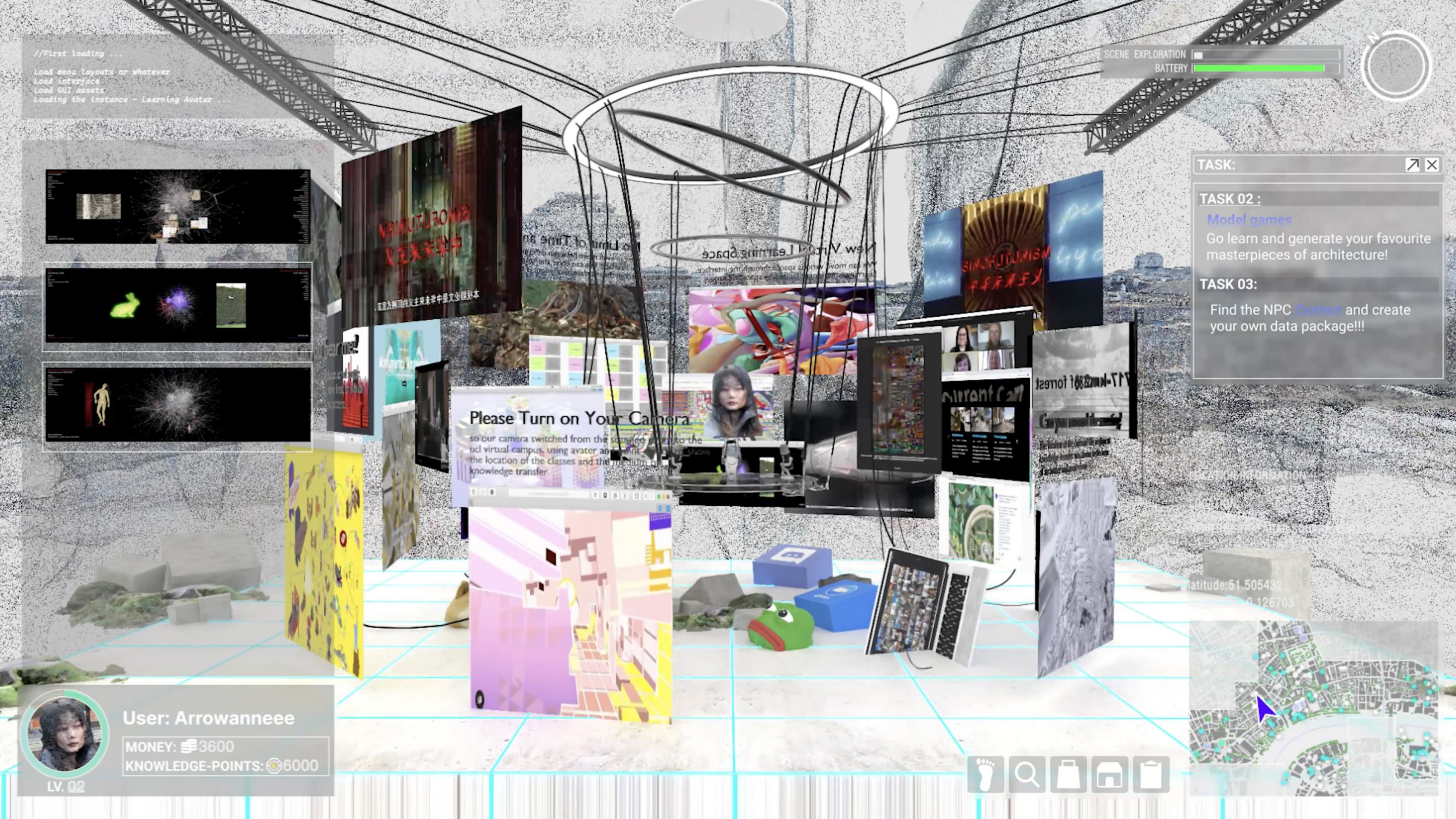}
        \label{fig:avatar2}
    \end{subfigure}
    \caption{An avatar as a posthuman learning agent situated between embodied perception, platform mediation, and urban infrastructure.}
    \label{fig:avatar}
\end{figure*}
The scene contracts from the urban and archival scale to a singular, centralized figure: a three-dimensionally scanned avatar positioned at the visual core of the system (Figure~\ref{fig:narrative}2; detail in Figure~\ref{fig:avatar}). Rendered with reflective surfaces, fragmented low-resolution textures, and visible computational noise, the avatar evokes the cold, machinic aesthetics characteristic of contemporary post-digital portraiture. Surrounding the figure, innumerable screens and interfaces proliferate, enclosing the avatar within a dense field of operational mediation. 

Within this configuration, the avatar functions not as a character or representational proxy, but as an infrastructural interface through which learning processes are enacted. The subject is no longer the origin or agent of knowledge acquisition; instead, it is reconstructed as a data object—scanned, modeled, and continuously processed by the learning system. Educational activity unfolds around and through this digitized body, positioning it as a node within a broader assemblage of data flows, interfaces, and algorithmic operations. 
Crucially, the scanned avatar stages a deliberate dislocation of subjectivity. Although it preserves the volumetric presence and surface details of a human portrait, it simultaneously exposes its technical condition through mesh fractures, uneven resolution, and residual scan artifacts. What confronts the viewer is not a reflection of the self, but an operationalized image—a body rendered legible to the system rather than expressive of interiority.

In this sense, the centralized avatar mirrors the logic of Mediatization Square: it does not reassert human subjectivity, but renders visible the mechanisms through which posthuman learning systems re-center cognition, identity, and agency as computable, inspectable, and infrastructural processes.


\subsubsection{The Dissipation of Institutional Space and Algorithmic Remnants} 
Educational institutions are no longer regarded as vessels for knowledge, but revealed as technical constructs that can be instantly invoked and revoked. Universities, museums and classical architecture appear as photogrammetric models, then dissolve into data, fragments, and other residual forms (Figure~\ref{fig:narrative}3). This concept comprises three visual narratives:
    \textbf{(1) Online virtual avatars congregating within a campus;} 
    \textbf{(2) Fragmented and dissipated the traditional institutional models;} 
    \textbf{(3) Classical column orders} showcase iconic architectural models in the digital age. Their photogrammetric models derive from image scraping of specific landmarks on social media platforms, resulting in highly uneven resolution across the same object: frequently photographed perspectives appear exceptionally sharp, while under-covered facets degrade into noise and gaps. This incompleteness reveals an algorithmic sacredness—classical forms no longer sustained by history, proportion, or symbolic systems, but shaped by platform attention, viewing frequency, and data density. Here, the sacred is recoded as the ``probability of being seen''.

By harnessing the digital technology of 3D modeling, the work embeds learning activities within a fracturable, dissipating geometric structure. It rejects the notion of posthuman learning as a ``decentralized state of freedom'', instead exposing it as a conditional reality shaped by computability, visibility, and the logic of interfaces.

\subsubsection{The Compressed City as Hyper-Learning Engine}
\begin{figure*}[h]
    \centering
    \begin{subfigure}[t]{0.9\textwidth}
        \centering
        \includegraphics[width=\linewidth]{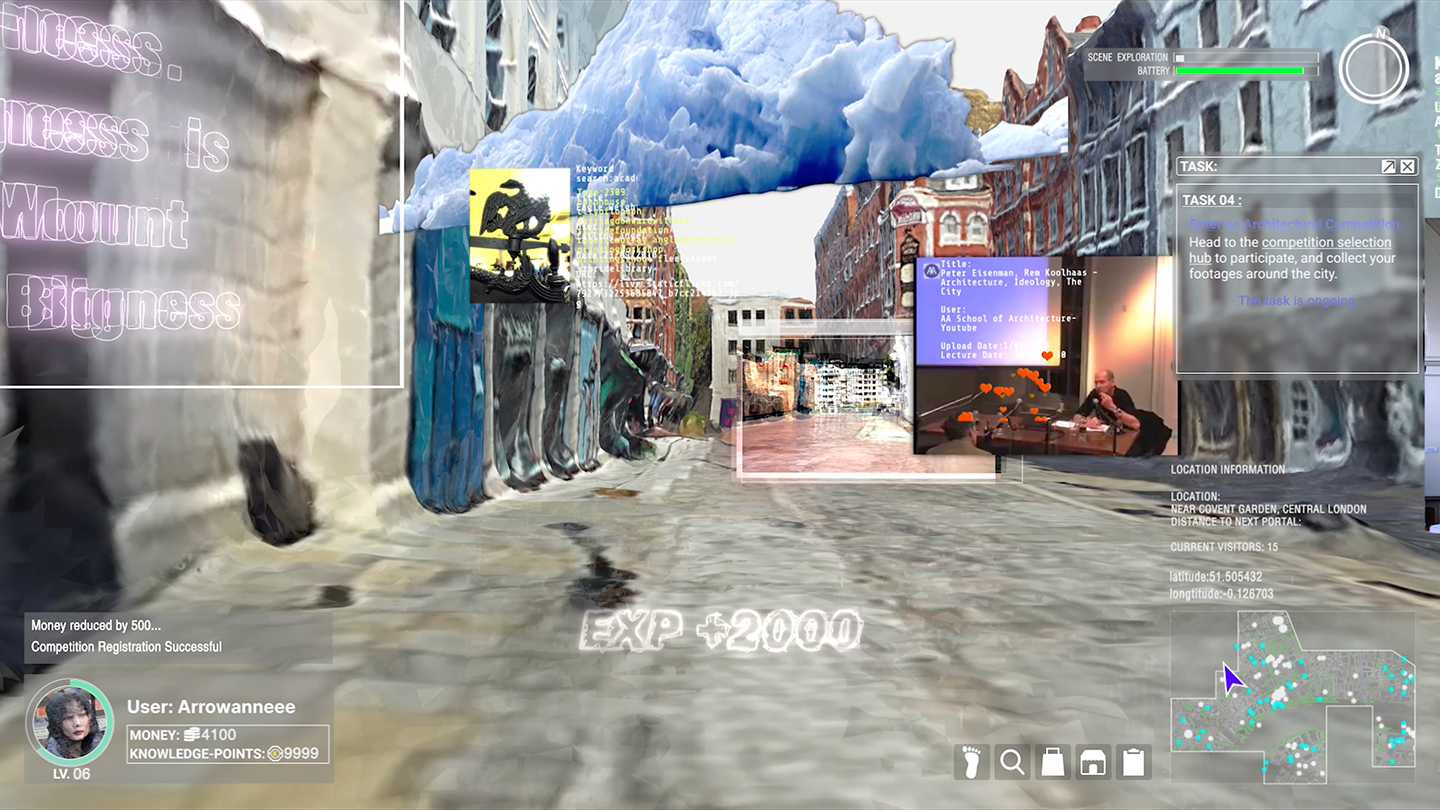}
        \label{fig:street1}
    \end{subfigure}
    \vspace{2mm}
    \begin{subfigure}[t]{0.9\textwidth}
        \centering
        \includegraphics[width=\linewidth]{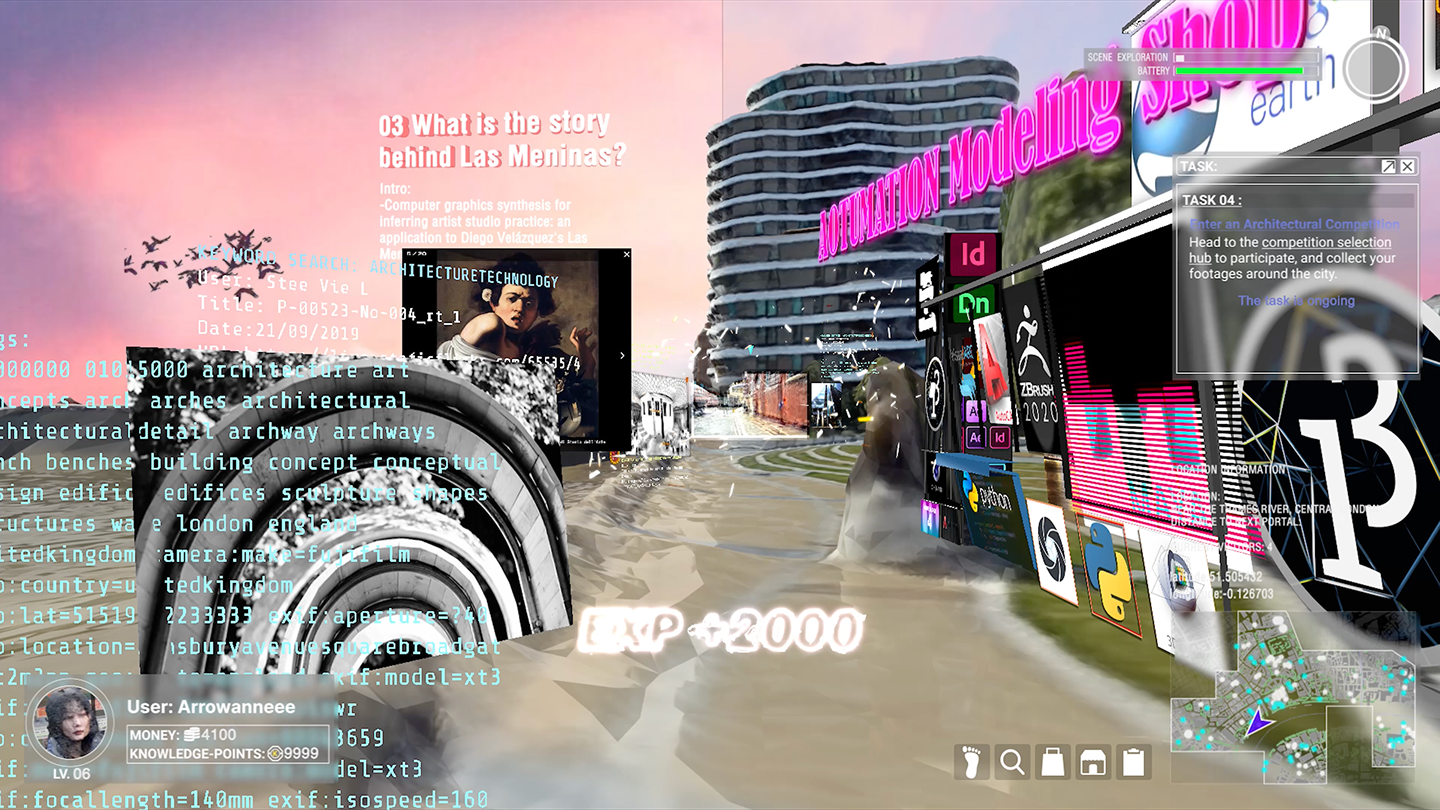}
        \label{fig:street2}
    \end{subfigure}

    \caption{City street scenarios within the XR environment, depicting situated encounters across layered media ecologies.}
    \label{fig:street}
\end{figure*}
As learning detaches from institutional walls, the city itself becomes a memory palace—an expanded cognitive architecture where photogrammetric fragments, user-generated traces, and algorithmic overlays assemble into a distributed archive (Figure~\ref{fig:narrative}F; details in Figure~\ref{fig:street}). Moving through London becomes a form of hyper-reading; each street corner is a knowledge node, each NPC a mediator of machine-curated wisdom, and each AR task a gesture of inscription into the collective memory of the compressed city.

\section{Content Generation Pipeline}
Based on this conceptual grounding, we produce this mock-up XR prototype using integrating multiple workflows to procedural animation, 3D modeling, materializing and rendering, and post-production effect. 


\paragraph{Narrative and Scenarios} 
\begin{figure}[h]
    \centering
    \begin{subfigure}[t]{\textwidth}
        \centering
        \includegraphics[height=0.48\textheight]{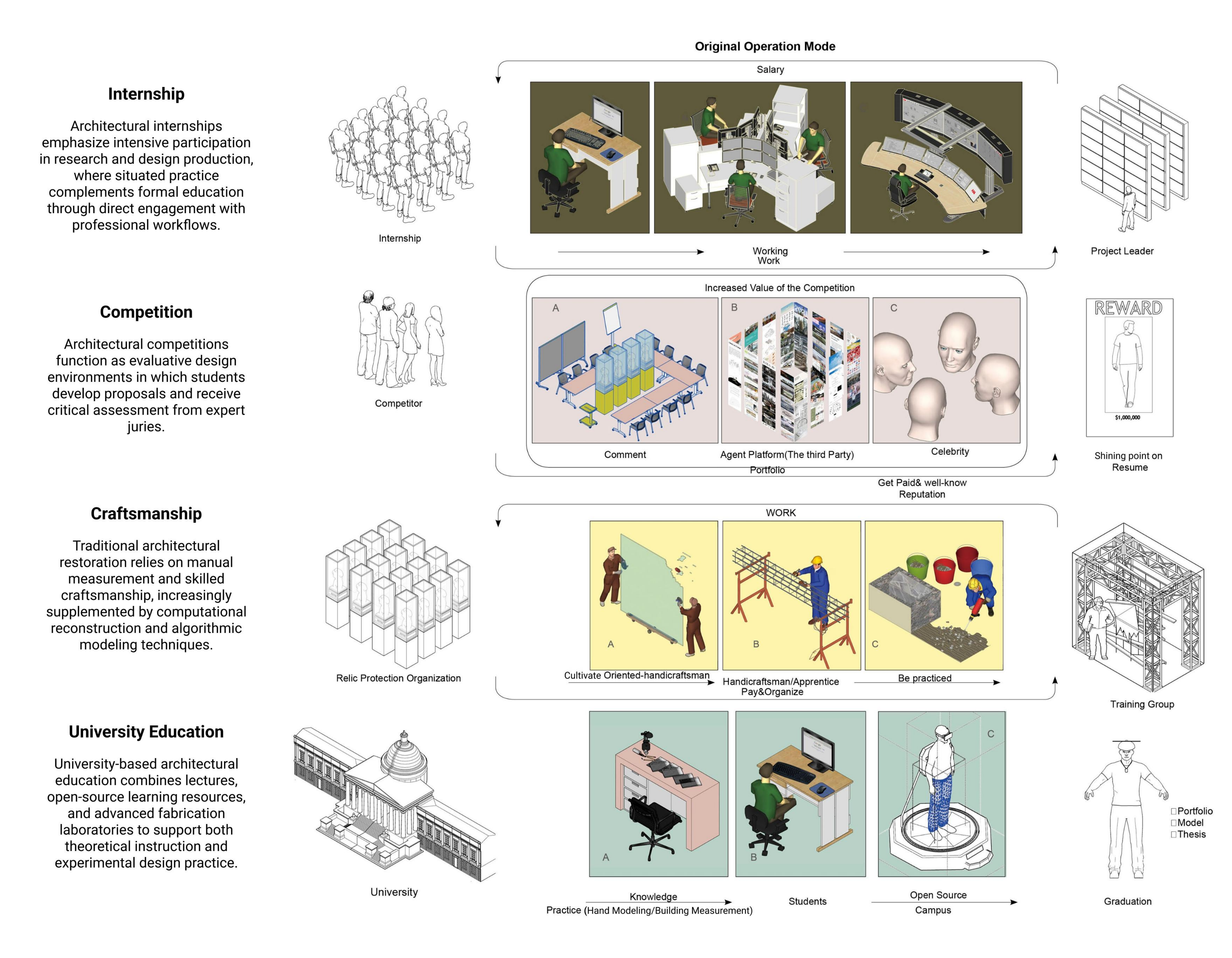}
        \label{fig:element1}
    \end{subfigure}
    \vspace{1mm}
    \begin{subfigure}[t]{\textwidth}
        \centering
        \includegraphics[height=0.38\textheight]{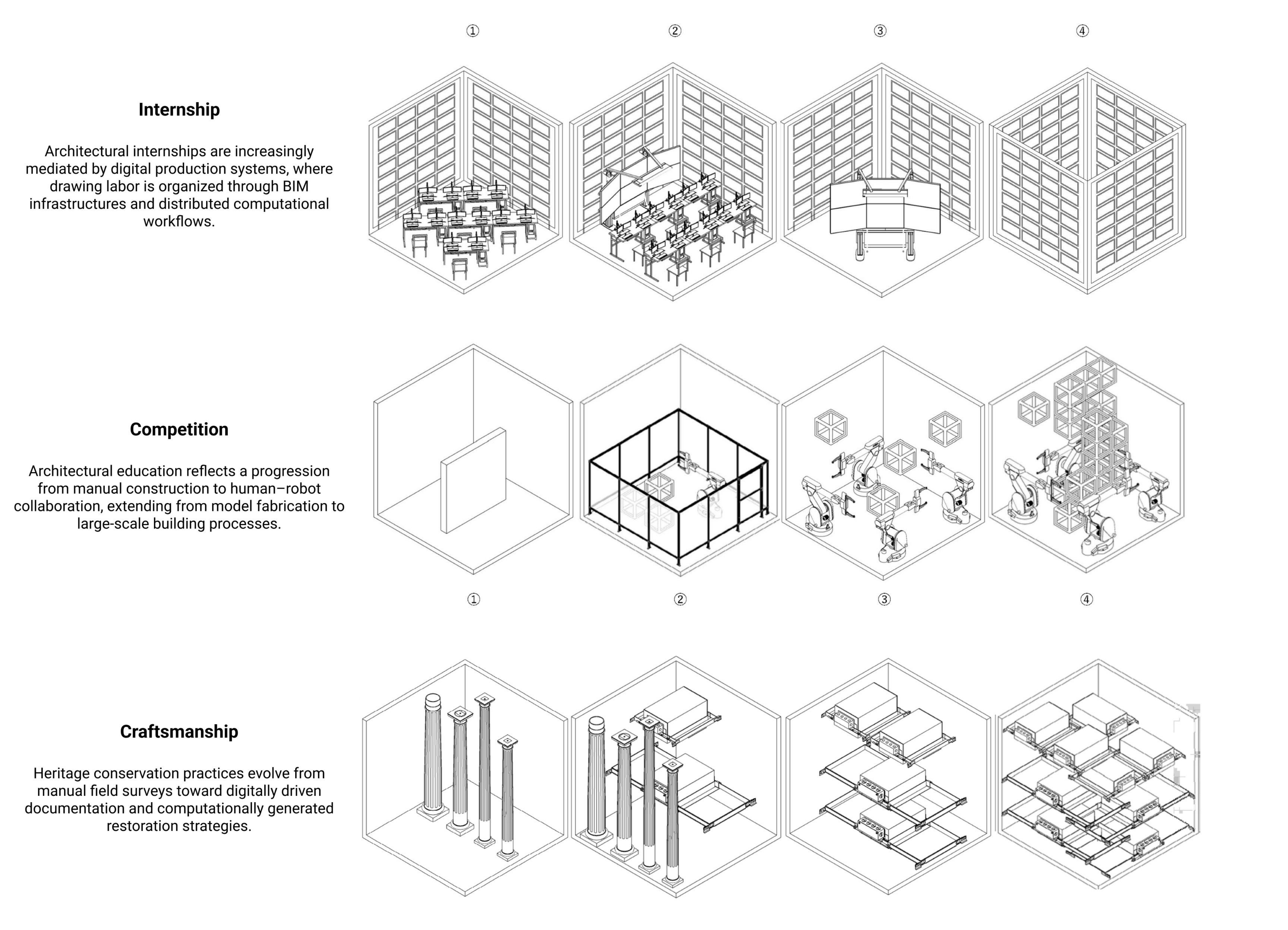}
        \label{fig:element2}
    \end{subfigure}
    \caption{One part of our narrative concept: Learning typology diagram juxtaposing distinct learning modalities in architectural education, reframed through a posthumanist perspective and curated as narrative components of the project.}
    \label{fig:narra_arch_elements}
\end{figure}
At the early stage of scenario production, we created storyboards to reflect the narrative structure. Through iterative discussions, we articulated core concept and elements (as Figure~\ref{fig:narra_arch_elements}), metaphors, and philosophical reflections associated with each lens. 
In parallel, we collected and developed visual aesthetic boards to guide the overall visual language and atmosphere of the mock-up XR experience. 

\paragraph{Media Collection and Scrapism} 
Social media images were incorporated as a primary data source reflecting the platform-mediated infrastructure of urban learning.  Publicly available Instagram photographs were collected using scenario-driven keywords and expanded through a snowball sampling strategy~\cite{biernacki1981snowball}. 
~
First, images were processed through creative coding workflows in Processing, where they were transformed into 3D semantic cloud structures. 
~
Second, a subset of images was employed as input for the further photogrammetric reconstruction, as described below. 

\paragraph{Spatial Reconstruction: Environment as Archive.} 
We constructed urban spaces, architectural elements, and situated objects that support narrative concepts and scenarios. 
Multiple acquisition strategies were combined to capture different spatial scales and levels of fidelity. (1) Large-scale urban contexts were reconstructed through 3D city data extracted from Google Maps using RenderDoc 1.9, (2) localized architectural and object-level details were generated through 3D scanning and photogrammetry. 

Mesh refinement, retopology, and material calibration were conducted in Blender, enabling photogrammetric models to be seamlessly integrated into the reconstructed environment. 
Camera placement and spatial framing were iteratively adjusted to align reconstructed environments with narrative viewpoints and intended user trajectories. 
The final environment assembly and spatial composition were completed using Blender, Houdini, and Rhino, enabling precise control over geometry, scale, and spatial relations across heterogeneous data sources.

\paragraph{Media-driven Procedural Transformation.} To support diverse narrative requirements and speculative representations, we developed a set of procedure-driven animation workflows that transform reconstructed models into dynamic visual artifacts. 
Using Processing and node-based visual programming in Houdini, we implemented procedural transformations including the conversion of 2D images into 3D point-cloud objects and the algorithmic manipulation of geometry and motion. 
By decoupling animation logic from manual keyframing, procedure-driven animation serves as a generative layer that mediates between captured reality and speculative narrative expression.

\section{Discussion}





This discussion situates the artwork across media ecology, posthuman theory, and planetary computation, examining how learning is reconfigured across institutional, technological, and planetary scales.

\subsection{Reflecting on Posthuman Agency and Technological Conditioning}
This project examines posthuman agency in learning systems thought with a weak form of technological determinism. Learning is framed not as an intentional human act, but as a conditioned process emerging from interactions among humans, platforms, architectures, and algorithmic operations~\cite{Winner1980}. Agency is therefore redistributed across heterogeneous actors, operating through platform logics that structure attention, memory, and action without fully determining outcomes.

Drawing on posthumanism's critique of human exceptionalism, the work foregrounds agency as relational and situational rather than autonomous~\cite{posthuman2013braidotti}. Unlike accounts that celebrate seamless human–machine symbiosis, this project emphasizes moments of negotiation, opacity, and partial loss of control. Learning unfolds through frictions between human intentions and computational procedures, revealing how algorithmic systems both enable and constrain epistemic agency~\cite{manovich2013}. 
By visualizing breakdowns, misalignments, and unresolved feedback loops, the project resists deterministic narratives in which technological systems appear as coherent or benevolent drivers of progress. Instead, technological conditioning is presented as uneven and contingent, producing shifting zones of dependence and uncertainty~\cite{mcluhan1964understanding}. Posthuman agency thus emerges not as empowerment alone, but as an ongoing negotiation within platform-governed learning environments, where cognition is continuously shaped, redirected, and destabilized by algorithmic mediation.

~~~~~~~~~~~~~~~~~~~~~~~~~~~~~~~~

\subsection{From Anthropocene to Planetary Epistemologies}
This work unveils that urban agency has already become planetary, while humanity is being forced to unlearn how to understand the systems within which we exist. \emph{Hyper-learning} belongs to cities, platforms, and models. \emph{Unlearning} indicates that human agency remains terrestrially embodied, epistemically outsourced, and structurally late. The tension between the two is not a future speculation—it is the present condition of learning in the compressed city. 

By doing this, the project reframes the concept of learning as a planetary process that exceeds human-centered epistemologies~\cite{Chakrabarty2009}. It rejects anthropocentric mastery by situating cognition within planetary media ecologies~\cite{Latour2017}. The artwork constructs a XR city as a planetary learning interface. Urban datasets, algorithms, and architectural systems co-perform cognition across human and nonhuman agents~\cite{bratton2015}. Learning emerges as an effect of infrastructural, ecological, and computational entanglements. Human knowledge is rendered contingent within geophysical, technical, and temporal scales~\cite{Crutzen2002}.

Unlike Anthropocene narratives that emphasize human geological dominance, planetary thinking suspends human centrality~\cite{Chakrabarty2021}. In contrast to Earth-system visualizations, this work stages lived epistemic disorientation within planetary computation. The project renders planetary abstraction experientially legible through architectural and media-based aesthetics. It shifts planetary theory from representational frameworks toward embodied epistemic negotiation~\cite{Latour2017}. For the Siggraph and broader art communities, the work expands computation beyond human-scale interaction. It positions planetary-scale media as a critical site for reimagining future learning imaginaries.

\section{Conclusion}
    \emph{Hyper-Learning \& Unlearning} reframes learning as a media-ecological condition shaped by the entanglement of urban space, computational systems, and posthuman infrastructures. Through a speculative XR city, the project exposes how learning is compressed across platforms, interfaces, and algorithmic processes that reorganize cognition, memory, and agency.

Rather than advancing pedagogical optimization, the work foregrounds learning as an infrastructural phenomenon embedded within software sovereignty. Educational institutions appear as recomposed residues—persisting through data circulation and computational visibility rather than institutional authority. In this context, \emph{Hyper-learning and Unlearning} operate together: accelerated access to knowledge coincides with the erosion of coherence and autonomy.

By situating agency within stacked relations among human bodies, machine intelligence, and urban environments, the project positions the city as a hyper-learning apparatus where action and understanding are conditioned by computational mediation. As an artistic intervention, \emph{Hyper-Learning \& Unlearning} offers an experiential critique of how learning and agency are reconfigured under planetary-scale computation.

\section{Acknowledgment} 
This project was developed as part of the graduation work for the UCL Bartlett B-Pro Urban Design Programme. The full project is available \href{https://drive.google.com/file/d/1-5bhx-WubRejsP46286mYmcm-HcHkNtc/view?usp=sharing}{here}. The authors would like to thank RC19 for their support throughout the project, and extend my sincere gratitude to the project tutors from \href{https://www.fieldstationstudio.org/}{Fieldstation Studio} — Corneel Cannaerts and Michiel Helbig — for their guidance. Special thanks also to the tutor Joris Putteneers, the theory tutor Provides Ng, and the workshop tutors James Melsom and Sam Lavigne for their valuable support and insights.

Additionally, the authors would like to acknowledge the use of the generative AI tool in this work. Specifically, \textit{ChatGPT-5.2} by OpenAI was utilized to assist in language refinement, including grammar and style corrections of existing manuscript text. All interpretations, conclusions, and final content remain the responsibility of the authors. 

\bibliographystyle{ACM-Reference-Format}
\bibliography{Hyperlearning}

\appendix

\end{document}